%% file: spie2018.tex
\documentclass[]{spie}  

\usepackage{amsmath,amsfonts,amssymb}
\usepackage{graphicx}
\usepackage[colorlinks=true, allcolors=blue]{hyperref}

\title{The Complicated Evolution of the ACIS Contamination Layer over the  
Mission Life of the Chandra X-ray Observatory }

\author[a]{Paul P. Plucinsky}
\author[a]{Akos Bogdan}
\author[b]{Herman L. Marshall}
\author[b]{Neil W. Tice}
\affil[a]{Harvard-Smithsonian Center for Astrophysics, MS-3, 60 Garden
St., Cambridge, MA 02138, USA}
\affil[b]{MIT Kavli Institute for Astrophysics and Space Research,
Cambridge, MA 02139, USA}

\authorinfo{Further author information: (Send correspondence to
  P.P.P)\\P.P.P.: E-mail: pplucinsky@cfa.harvard.edu}

\newcommand{\etal}{{\em et al.}}

\DeclareRobustCommand{\ion}[2]{\textup{#1\,\textsc{\lowercase{#2}}}}
 
\begin{document} 
\maketitle

\begin{abstract}

  The {\it Chandra X-ray Observatory} (CXO) was launched almost 19 years ago and has been 
delivering spectacular science over the course of its mission.  The 
{\it Advanced CCD Imager Spectrometer} (ACIS) is the prime instrument on the satellite, 
conducting over 90\% of the observations. The CCDs operate at a temperature
of -120 C and the optical blocking filter (OBF) in front of the CCDs is at a
temperature of approximately $-60$~C.  The surface of the OBF has 
accumulated
a layer of contamination over the course of the mission, as it is the coldest
surface exposed to the interior to the spacecraft.  We have been 
characterizing the 
thickness, chemical composition, and spatial distribution of the contamination
layer as a function of time over the mission. All three have exhibited 
significant changes with time. There has been a dramatic decrease in
the accumulation rate of the contaminant starting in 2017.  The
lower accumulation rate may be due to a decrease in the deposition
rate or an increase in the vaporization rate or a combination of the two. 
We show that the current calibration file which models the additional
absorption of the contamination layer is significantly overestimating
that additional absorption by using the 
standard model spectrum for the supernova remnant 1E~0102.2-7219
developed by the {\em International Astronomical Consortium for High Energy 
Calibration} (IACHEC).  In addition, spectral data from the cluster of
galaxies known as Abell~1795 and the Blazar Markarian 421 are used to
generate a model of the absorption produced by the contamination
layer.  The {\it Chandra X-ray Center} (CXC) calibration team is
preparing a revised calibration file that more accurately represents
the complex time dependence of the accumulation rate, the spatial
dependence, and the chemical composition of the contaminant.  Given
the rapid changes in
the contamination layer over the past year, future calibration
observations at a higher cadence will be necessary to more accurately
monitor such changes.

\end{abstract}


\section{INTRODUCTION}
\label{sec:intro}  

The {\it Chandra X-ray Observatory} (CXO) was launched on 23 July 1999
on the Space Shuttle {\it Columbia}.  An overview of the mission and
its instruments are presented in 
Weisskopf~\etal~(2000)~\cite{weisskopf2000} and an update on
the mission was provided in Weisskopf~\etal~(2012)\cite{weisskopf2012}.  
The CXO carries two imaging instruments, the {\it Advanced CCD Imaging
Spectrometer} (ACIS) discussed in Garmire~\etal~(1992)~\cite{garmire92}
and Garmire~\etal~(2003)~\cite{garmire03} and the {\it High
Resolution Camera} (HRC) discussed in
Murray~\etal~(1997)~\cite{murray1997}.  
In addition, the CXO
carries two gratings instruments known as the {\it High Energy
  Transmission Grating} (HETG) described in 
Canizares~\etal~(2000)~\cite{canizares2000} and the {\it Low
  Energy Transmissions Grating} (LETG) 
Brinkman~\etal~(2000)~\cite{brinkman2000}.

ACIS is the primary instrument on the CXO with a nominal bandpass of
 0.3--10.0~keV, conducting over 90\% of the
observations.  ACIS contains 10 CCDs arranged into two arrays. One
array, the ACIS Imaging array (ACIS-I), consists of four frontside
illuminated (FI) CCDs arranged in a $2\times2$ array, and the other
array, the ACIS Spectroscopy array (ACIS-S), consists of  four FI CCDs
and two backside illuminated (BI) CCDs arranged
in a $1\times6$ array. The ACIS-I array is used primarily for imaging
spectroscopy and the ACIS-S array is used primarily as the readout
detector for the HETG and LETG, although the ACIS-S is also used for
imaging spectroscopy.  The BI CCDs have higher quantum efficiency at
low energies than the FI CCDs and are therefore preferred over the FI
CCDs for some imaging observations.

 In order to suppress optical to infrared photons but to transmit the
 X-ray photons of interest, both ACIS arrays have an {\em Optical
   Blocking Filter} (OBF) inserted in the optical path.  The filters 
were produced by ${\tt
   Luxel^{TM}}$ and are made of polyimide with Al deposited on both
 sides of the polyimide.  The two filters are of slightly different
 thicknesses, the ACIS-S OBF (OBF-S) is 100/200/30 nm of
 Al/polyimide/Al and the ACIS-I OBF is 130/200/30 nm of
 Al/polyimide/Al. The OBFs sit about 12~mm in front of the CCDs facing
 the mirrors on the CXO.  The volume around the CCDs is effectively
 isolated from the interior of the spacecraft, while the surface of
 the OBFs facing the mirrors is exposed to the interior of the
 spacecraft. The CCD focal plane is regulated at a temperature of
 -120~C.  The OBFs are positioned at the top of the ACIS Camera Body (CB)
 which was regulated at -60~C early in the mission, but has been
 unregulated from April 2008 fluctuatng between between -72~C and
 -60~C. The CB was regulated at -60~C from August 2015 until July 2016
 but has since been unregulated (see
 Plucinsky~\etal~2016~\cite{plucinsky2016}  for details).
 In normal operations, the centers of the
 filters are warmer by $\sim2-4$ degrees due to the radiative heat load
 of the warm mirrors (+20~C) and the optical bench assembly.

 It was noticed early in the mission\cite{plucinsky2003} that the low
 energy sensitivity of the ACIS instrument was decreasing with time.
 It was quickly determined that this loss of detection efficiency was
 the result of a contamination layer building up on the surface of the
 OBFs facing the spacecraft interior.  The contamination layer
 continues to accumulate even after 18 years on orbit.  The
 accumulation rate, the chemical composition, and the spatial
 distribution of the contaminant have all varied with time over the
 mission. The accumulation rate exhibited a steep rise at the
 beginning of the mission, a flattening from 2003 until 2010, and
 then another steep rise from 2010 onwards. We reported in 
 2016\cite{plucinsky2016} on our efforts to reduce the accumulation
 rate by turning on the ACIS Detector Housing (DH) heater which regulates
 the CB and hence OBF edges at -60~C.  There was no
 measurable effect on the accumulation rate due to the DH heater
 regulating the CB at -60~C.  In this paper we report that the
 accumulation rate has decreased significantly starting in 2017
 and we discuss our current understanding of the
 time-variable accumulation rate and chemical composition.

\section{ACIS Contamination Layer} 
\label{sec:contam}

\subsection{Discovery and Initial Characterization} 
\label{sec:discovery}

The existence of the contamination layer was discovered in 
2002\cite{plucinsky2003} as a gradual decrease in the  low energy 
detection efficiency of all of the CCDs. The growth of
the contamination layer was tracked by repeated observations of the
{\em external calibration source}~(ECS) which has lines of Al-K~(1.5
keV), Ti-K~(4.5~keV), and Mn-K~(5.9~keV) from an ${\rm Fe^{55}}$
radioactive source with a half-life of 2.7~yr. The ECS also produced a
line complex from Mn-L around 0.67 keV.  The ratio of the Mn-L/Mn-K
count rates
on the S3 BI CCD became the most useful measure of the declining
sensitivity at low energies. Unfortunately, the observed flux from the
Mn-L line complex decreased with time due to the decay of the
radioactive source and the increasing thickness of the contamination
layer.  Eventually the uncertainties on the measurements became so
large that they were no longer useful to track the growth of the
contamination layer. As
the mission progressed, we transitioned to using celestial sources to
monitor the growth of the contamination layer. We used celestial
sources that are believed to be constant (or nearly constant on human
time scales), such as clusters of galaxies and supernova remnants
(SNRs), to monitor the change in low energy detection efficiency.  We
also used bright, variable sources with the HETG and LETG to constrain
the absorption as a function of energy produced by the contaminant.

Early efforts to determine the chemical composition of the 
contaminant\cite{marshall2004} identified absorption edges of C, O,
and F that were in excess of the edges in the ACIS OBF.  The ACIS OBF
has absorption edges of C and O, but no edge due to F. The ACIS
detection efficiency as a function of energy was carefully calibrated
before launch\cite{bautz1998,nousek1998} including the transmission and 
absorption edges of the OBFs.  The flight measurements used a bright 
continuum source dispersed with the HETG and/or LETG to achieve the 
highest spectral resolution possible with the CXO.  In these high 
resolution spectra, it became obvious that some absorption edges were
deeper than in the pre-flight measurements or only appeared (in the
case of F) after launch.  The newly-detected absorption edges were
also found to be increasing in time. C was by far the dominant species
in the contaminant while the O and F were approximately equal in
concentration.  We believe the contaminant started accumulating as
soon as the ACIS door was opened and the OBFs were exposed to the
interior of the spacecraft.  The contaminant has continued to
accumulate for the entire 19 year mission of the CXO, see 
Section~\ref{sec:accumulation} for a detailed time history.

\subsection{OBF and Camera Body Temperatures} 
\label{sec:heater}

  ACIS has two separate filters, one for the Imaging array, OBF-I, and
  one for the spectroscopy array, OBF-S. For diagrams and pictures of
  the flight hardware, see the figures in
  Plucinsky~\etal~(2004)\cite{plucinsky2004}.  Both OBFs are secured to
  the top surface of the ACIS Camera Body (CB).  The OBFs have no active
  thermal control but respond to the environment around them. The
  edges of the filter are in good thermal contact with the CB and are
  therefore at the same temperature as the CB. At the beginning of the
  mission, the CB was held at -60~C.  The centers of the
  filter are warmer than the edges due to the radiative heat load from
  the warm mirrors and optical bench cavity. The center of the OBF-I
  is modeled to achieve a temperature of $\sim-56$~C while the center
  of the OBF-S is at  about $\sim-58$~C. There is no temperature
  sensor on the OBFs themselves.

  In April 2008, it was decided to turn off the ACIS Detector Housing
  (DH) heater which kept the CB temperature at -60 C.  With the DH
  heater off, the CB temperature fluctuated between  -72~C and -62.5~C
  depending on the orientation of the CXO spacecraft.  The cooler CB
  temperature provided more margin for keeping the CCDs in the focal
  plane at -120~C.  
  From
  launch in 1999 until April 2008, the CB temperature regulated at
  -59.9~C except for a few excursions during special activities.
  After April 2008, the CB temperature was unregulated and varied
  with the orientation of the spacecraft.  
   In August 2015, it was decided to turn the
  ACIS DH heater back on with the hope that the accumulation rate of
  the contaminant would decrease.  But as we reported in
  2016\cite{plucinsky2016}, the warmer CB temperatures has no effect
  on the accumulation rate of the contaminant.  Therefore, it was
  decided in July 2016 to turn the DH heater back off and leave the
  CB temperatures unregulated.

   \begin{figure}
   \begin{center}
   \begin{tabular}{c}
   \includegraphics{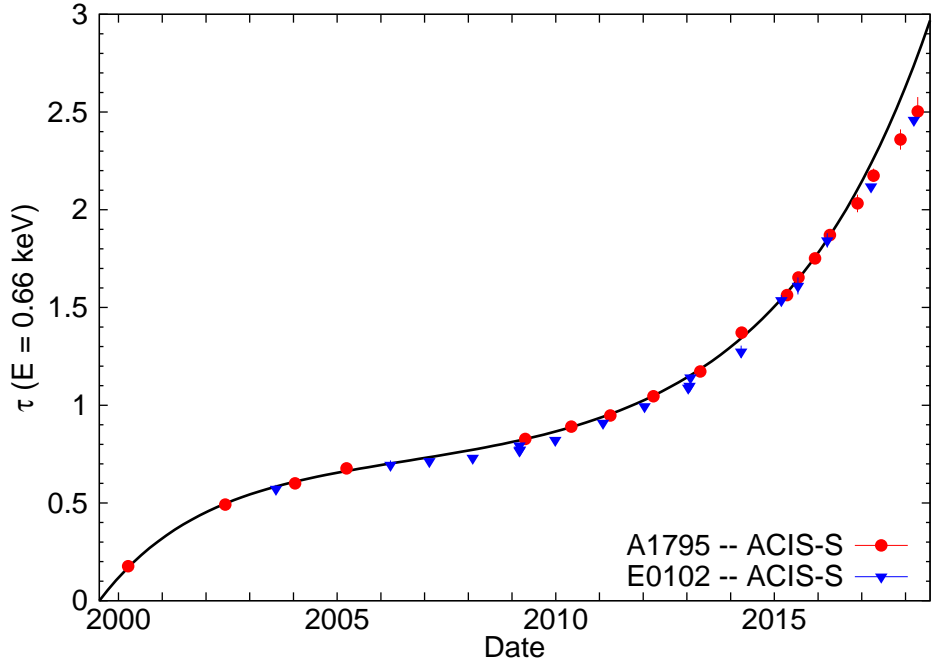}
   \end{tabular}
   \end{center}
   \caption[]
   { \label{fig:akos_aciss} Optical depth at 0.67 keV for the ACIS-S
     aimpoint as
     determined by fits to the E0102(blue) and A1795(red) data.   The
     black line is the model for the optical depth in the
     N0010 contamination model.

}
  \end{figure} 

\subsection{Time Dependence of the Accumulation Rate} 
\label{sec:accumulation}

As mentioned in Section~\ref{sec:discovery}, the accumulation rate of
the contamination layer was monitored with the ECS until the
radioactive source became too faint to produce reliable results.  At
this point, we switched to using the brightest SNR in the Small Magellanic
Cloud, 1E~0102.2-7219 (hereafter E0102), a bright cluster of galaxies
known as Abell~1795 (hereafter A1795), and a bright Blazar called 
Markarian 421 (hereafter Mkn~421). E0102 has a soft, line-dominated
spectrum and we have used it throughout the mission to characterize the
contamination layer\cite{plucinsky2008,plucinsky2012}. The development
of the standard IACHEC model for E0102 and its application to the
current generation of X-ray instruments is presented in our 2017
paper\cite{plucinsky2017}.  A1795 has a
harder thermal spectrum with some significant line emission.  Mkn~421
has a continuum spectrum described by a curved power-law model.

   \begin{figure}
   \begin{center}
   \begin{tabular}{c}
   \includegraphics{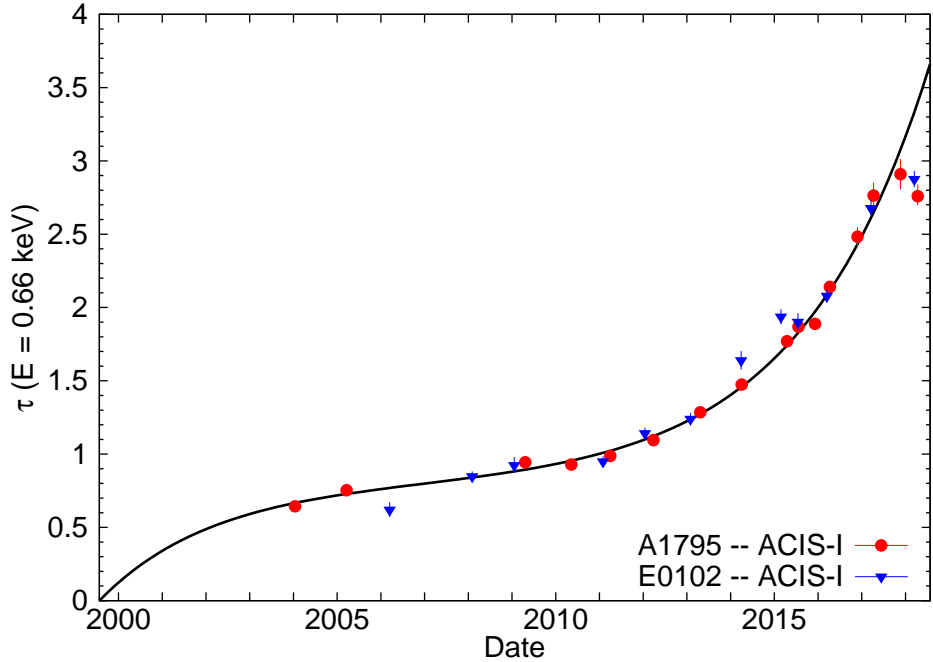}
   \end{tabular}
   \end{center}
   \caption[]
   { \label{fig:akos_acisi} Optical depth at 0.67 keV for the ACIS-I
     aimpoint as
     determined by fits to the E0102(blue) and A1795(red) data.   The
     black line is the model for the optical depth in the
     N0010 contamination model.

}
  \end{figure} 

\begin{figure}
   \begin{center}
   \begin{tabular}{c}
   \includegraphics{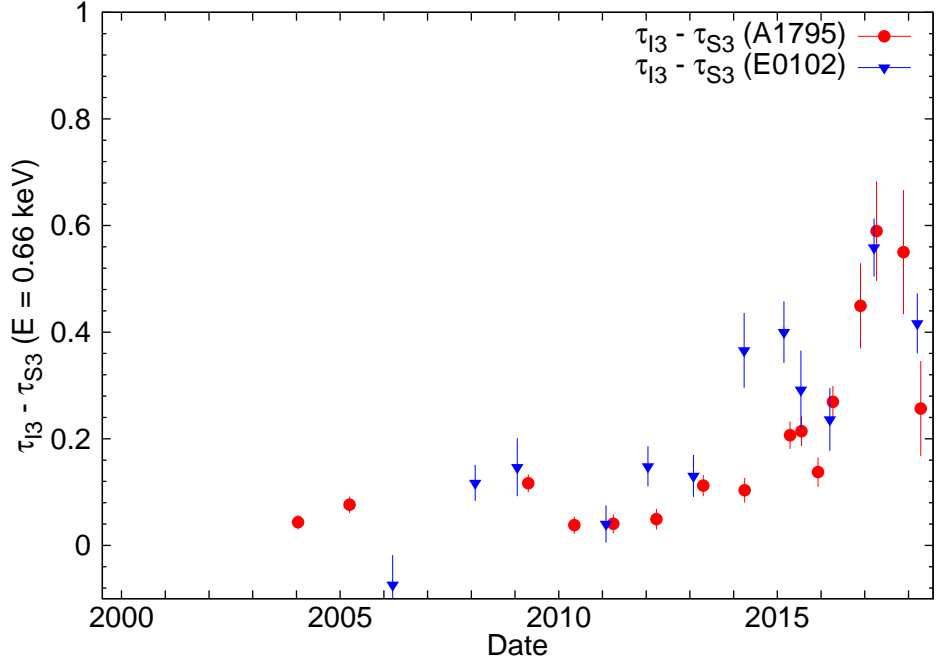}
   \end{tabular}
   \end{center}
   \caption[]
   { \label{fig:od_diff} The difference in the optical depth at 0.67
     keV at the ACIS-I and ACIS-S aimpoints.  The contaminant grew
     more quickly on OBF-I than OBF-S from 2015 until 2017.

}
  \end{figure}

We have used the A1795 and E0102 data on ACIS-S and ACIS-I to measure the
optical depth of the contaminant at 0.67~keV (the energy of the Mn-L
complex in the ECS).  The results for the ACIS-S aimpoint are plotted in 
Figure~\ref{fig:akos_aciss}.  The blue data points are derived from
the E0102 data, the red data points are derived from the A1795 data
and the black curve shows the expected increase in the contamination 
layer that is contained in the current release of the CXC
contamination file ``{\tt acisD1999-08-13contamN0010.fits}'', called
``N0010'' for short. The measured optical depths from
the E0102 and A1795 data are consistent within the uncertainties.
The accumulation history of the contaminant is 
shown in this figure, a steep rise early in the mission, a reduction
in the rate from 2003 to 2010, another sharp increase after 2010, and
an apparent decrease starting in 2017. The data in 2017 begin to deviate
from the expected accumulation rate and the trend continues into
2018.  The decrease in the accumulation rate
is not correlated with the DH heater which was on from 11 August 2015
until 20 July 2016. The behavior at the aimpoint on ACIS-I is even
more dramatic and shown in Figure~\ref{fig:akos_acisi}.  The first
data point to deviate from the expectation is in late 2017.  Perhaps
more interesting, the last data point in 2018 is consistent with no
accumulation over the last 6 months. The uncertainties are relatively
large so future measurements will be necessary to confirm this
result. Note that the maximum optical depth is about 3.0 on the OBF-I
and is about 2.5 on the OBF-S.  The contaminant has apparently
accumulated more rapidly at the center of the OBF-I than at the center
of the OBF-S.  This can be seen more clearly in
Figure~\ref{fig:od_diff} which shows the difference in the optical
depth at the aimpoints on ACIS-I and ACIS-S as a function of time.
For most of the mission,
the optical depths were within 0.2 of each other.  But starting in
2015, the contaminant grew more rapidly near the aimpoint on ACIS-I
reaching a maximum difference of 0.6 optical depths.  Curiously, the
most recent data point in 2018 shows the difference between OBF-I and
OBF-S is decreasing. This suggests that the accumulation rate on OBF-I
is close to zero while the accumulation rate is still positive and
small on  OBF-S.
Future observations will be necessary to determine if this trend will 
continue.

%

\subsection{Time Dependence of the Chemical Composition} 
\label{sec:composition}

\begin{figure}
   \begin{center}
   \begin{tabular}{c}
   \includegraphics[width=2.4in,angle=90]{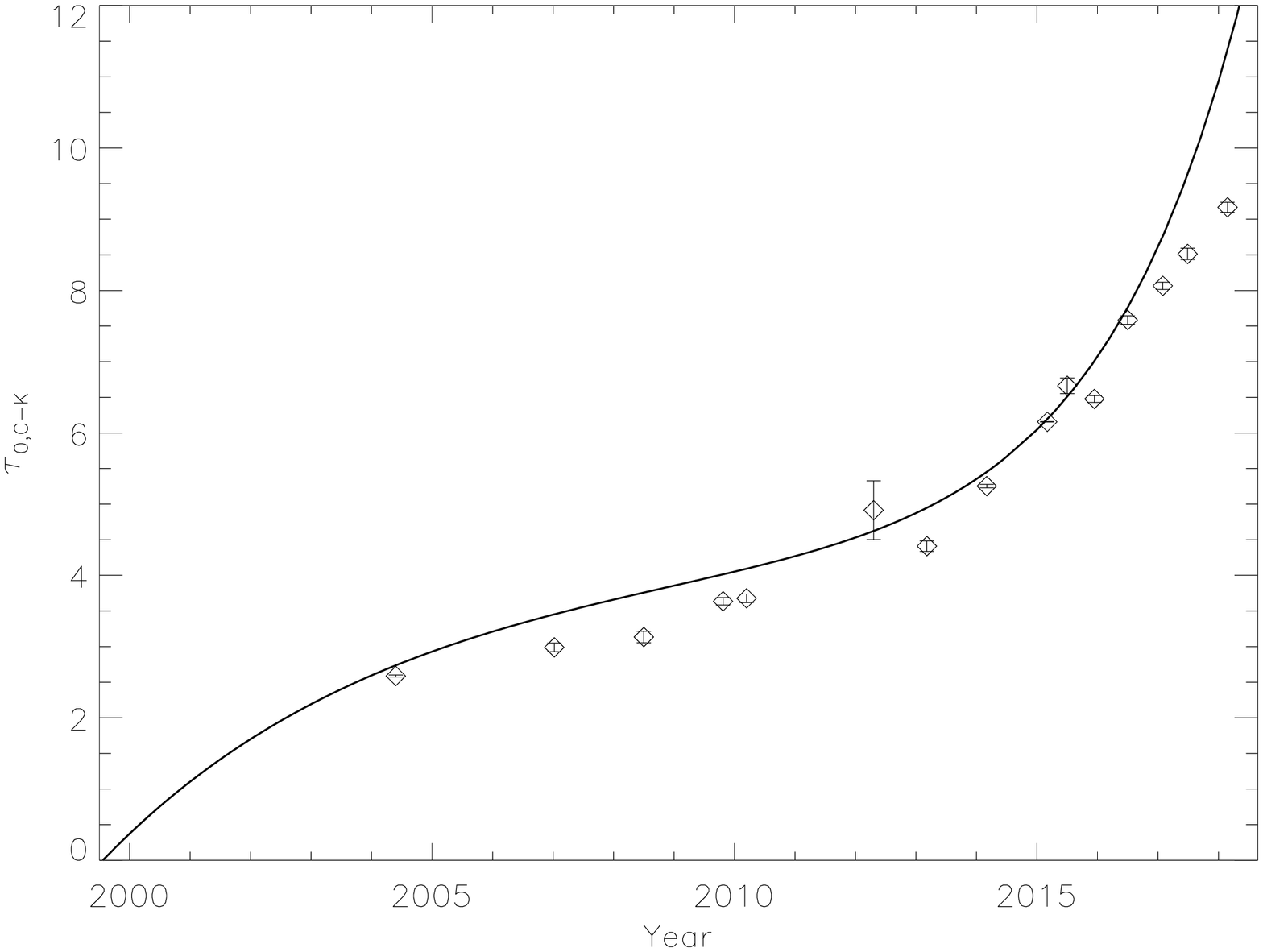}
   \includegraphics[width=2.4in,angle=90]{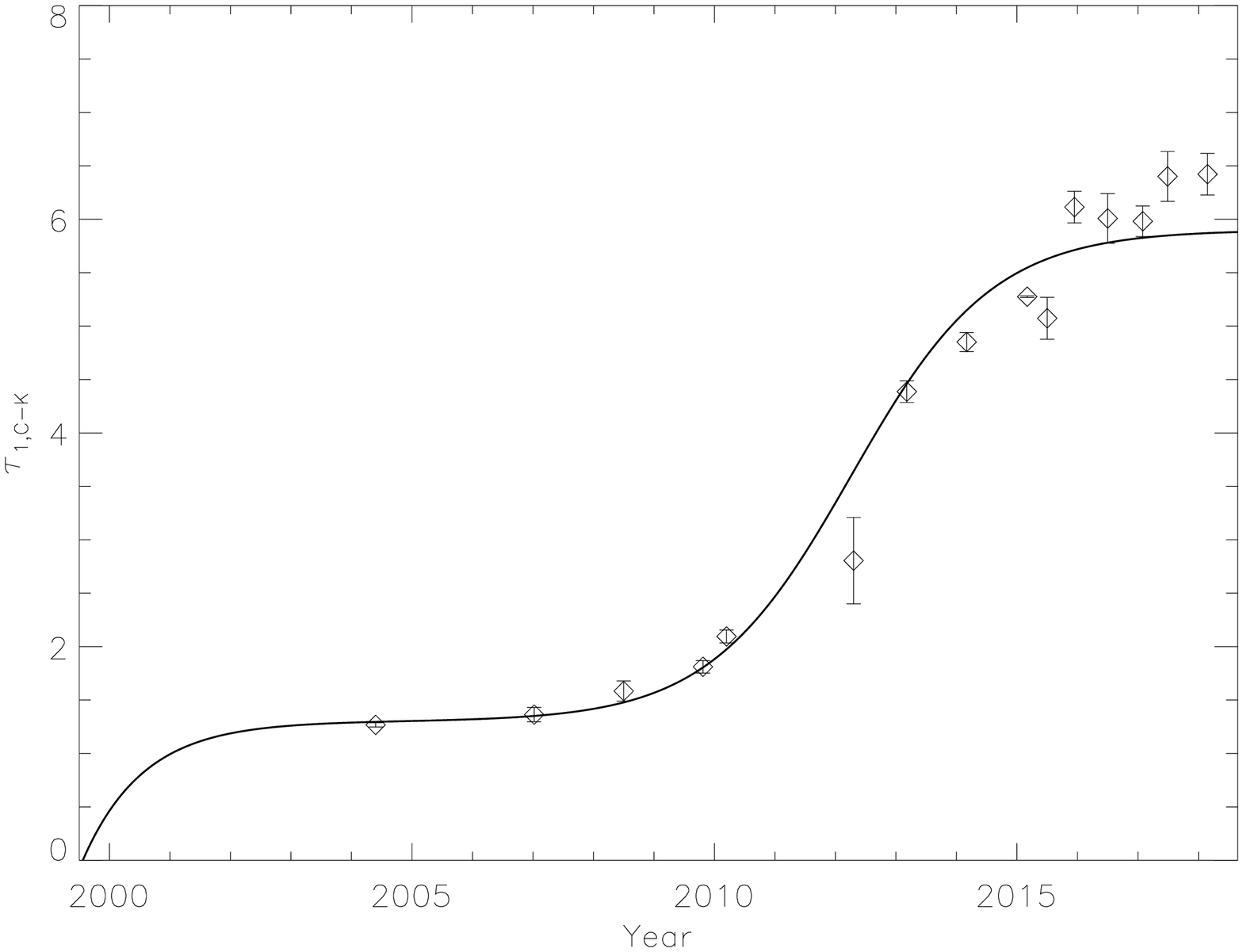}
   \end{tabular}
   \end{center}
   \caption[]
   { \label{fig:ck_tau0} LEFT: The optical depth at C-K of the spatially
uniform component near the center of the ACIS-S array. RIGHT: The
optical depth at C-K of the spatially
variable component near the center of the ACIS-S array.
The solid line for both is the prediction from the N0010 contamination
model.
}
   \end{figure}

The high resolution spectra provided by the HETG have been 
used\cite{marshall2004} to constrain the chemical composition of the 
contaminant and how it has changed with time.  The contaminant is composed
mostly of C, with some O and F. One limitation of the HETG data is
that they only provide information on the contaminant for the OBF-S
filter. The optical depth of the contaminant
for each element (C, O, \& F) is modeled as a functions of time and
position with a two component model:

\begin{center}

$\tau(t;x,y) = \tau_0(t) + [\tau_1(t) \times f(x,y)]$

\end{center}

\noindent where $\tau_0(t)$ represents a time-variable, spatially uniform component,
$\tau_1(t)$ represents a time-variable, spatially variable component,
and $ f(x,y)$ is the spatial distribution for the spatially variable
component. Figure~\ref{fig:ck_tau0} shows the time dependence of the
 $\tau_0(t)$ and  $\tau_1(t)$ components for C near the aimpoint on
 the ACIS-S detector derived from HETG observations of Mkn~421.  The 
time dependence of $\tau_0(t)$ for C  matches that of the N0010 model
until the last few data points which are significantly
below the line. This is similar to the behavior seen for A1795 and
E0102 shown in Figure~\ref{fig:akos_aciss}.
  The  time dependence of $\tau_1(t)$ for C matches
that of the N0010 model in shape, but the N0010 model might be
slightly under-predicting at late times.  The  $\tau_1(t)$ component
has been mostly flat
with time from 2015 onwards, while the  $\tau_0(t)$ continues to
accumulate, albeit at a lower rate than predicted by the N0010 model.
One interpretation of this behavior is that the spatially uniform 
component and the spatially variable component correspond to separate
materials and the spatially variable component has ceased to
accumulate.

\begin{figure}
   \begin{center}
   \begin{tabular}{c}
   \includegraphics[width=2.4in,angle=90]{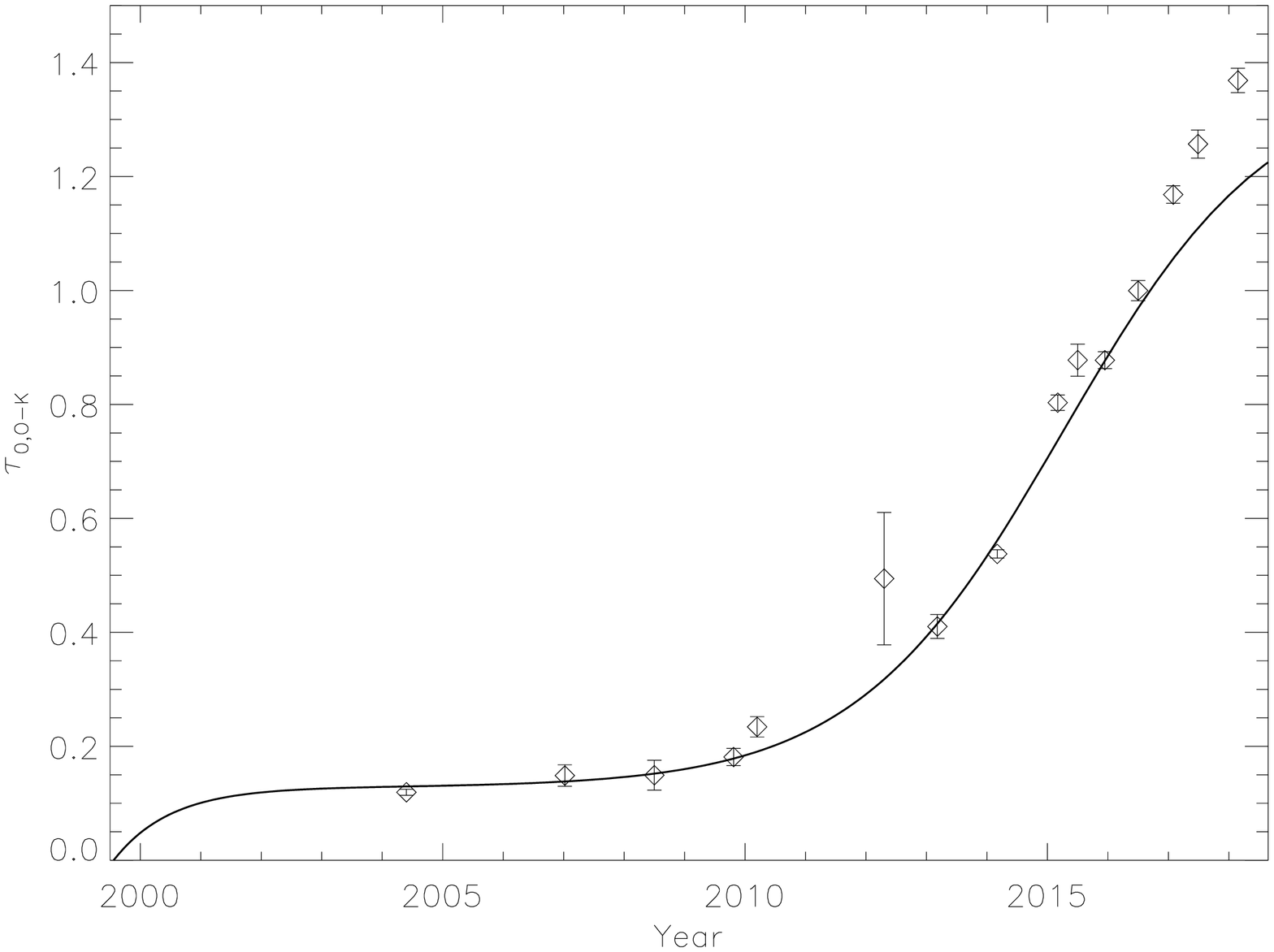}
   \includegraphics[width=2.4in,angle=90]{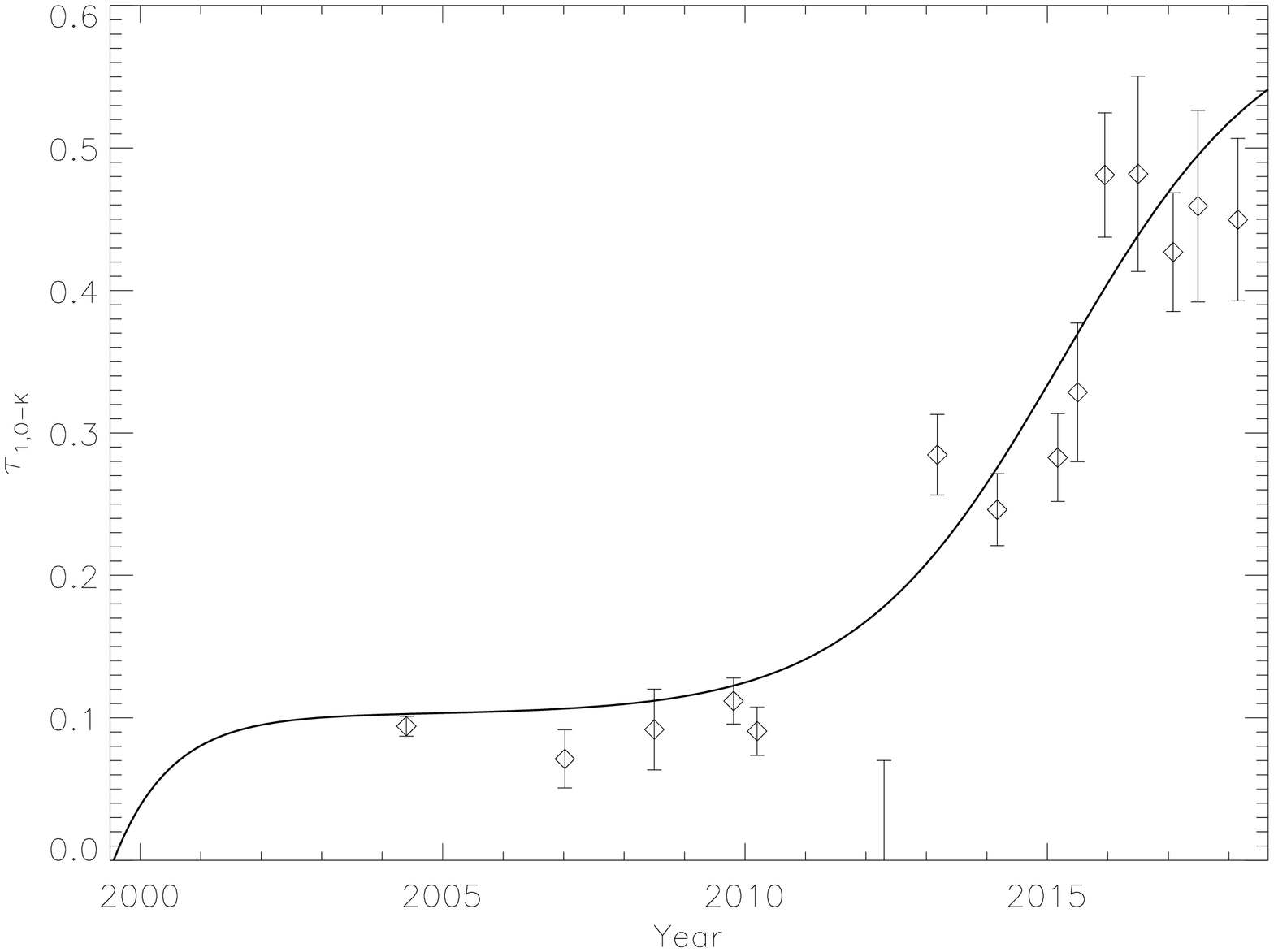}
   \end{tabular}
   \end{center}
   \caption[]
   { \label{fig:ok_tau0} LEFT: The optical depth at O-K of the spatially
uniform component near the center of the ACIS-S array. RIGHT:
The optical depth at O-K of the spatially
variable component near the center of the ACIS-S array.
The solid line for both is the prediction from the N0010 contamination
model.
}
   \end{figure}

Figure~\ref{fig:ok_tau0} shows the time dependence  $\tau_0(t)$ and
$\tau_1(t)$ components for O, again near the aimpoint on the ACIS-S
detector.  The time dependence of $\tau_0(t)$ for O  matches that of
the N0010 model until the last few data points
which are significantly above the line.  
   The time dependence 
of $\tau_1(t)$ for O matches that of the N0010 model in shape and
amplitude.  However, the data since 2015 are consistent with no growth in this
component so the N0010 model may be over-predicting the contaminant at
late times but the uncertainties are still large enough that the case
is not definitive.
The  $\tau_0(t)$ result indicates that the N0010
model has less O than it should.  But note that the total optical
depth of O is significantly less than that of C, $\sim2.0$ versus
$\sim15$, so that any error in the O optical depth has less effect on
the observed spectra and is therefore more difficult to discern.

%

%

\section{Performance of the Current Contamination File} 
\label{sec:verification}

\subsection{CXC Calibration Files} 
\label{sec:calibration_files}

 The CXC calibration group is responsible for providing calibration
 files that accurately model the additional absorption produced by the
 contamination layer.  As mentioned above the characterization of the
 contamination layer is complicated by the temporal variation of the
 thickness, the chemical composition and the spatial distribution. The
 CXC regularly acquires calibration data of standard targets such as
 E0102, A1795, and Mkn~421 to verify the current contamination
 calibration file. If deficiencies are found, a new calibration file
 is created to address those deficiencies.  The ACIS contamination
 file has been updated 7 times over the course of the CXO mission. For
 the analysis that follows, we use version {\tt N0010} of the model,
 which is called {\tt acisD1999-08-13contamN0010.fits} in the CXC
 {\em Calibration Database}~(CALDB).  We used {\em Chandra Interactive
   Analysis of Observations}~(CIAO) version~4.9 and CALDB version~4.7.8.
 
\subsection{E0102 Model}
\label{sec:e0102_model}

 We have defined a standard model for E0102 as part of the activities
 of the IACHEC.  We have used this model
 extensively~\cite{plucinsky2008,plucinsky2012,plucinsky2017} to test and improve
 the ACIS response model earlier in the mission. The model is
 intended for calibration analyses and is not intended to provide any
 insight into E0102 as a SNR.  The model is
 empirical in that it uses 52 Gaussians to model the line emission.  
It uses a two component absorption model, one component for the
Galactic contribution and one for the Small Magellanic Cloud (SMC) contribution.  
We modeled the continuum using a modified version of the
{\em Astrophysical Plasma Emission Code}~({\tt
  APEC})~\cite{smith2001} called the {\tt ``No-Line''} model.   
This model excludes all line emission, while retaining all continuum  
processes including bremsstrahlung, radiative recombination continua  
(RRC), and the two-photon continuum from hydrogenic and helium-like  
ions (from the strictly forbidden ${}^2S_{1/2} 2s \rightarrow$\ gnd  
and ${}^1S_0 1s2s \rightarrow$\ gnd transitions, respectively). 
We included two continuum components of this type in the E0102 model.
For details of the model and the parameters assumed see
Plucinsky~\etal~(2017)\cite{plucinsky2017}.

\input{s3tab.tex}

 Although the standard IACHEC model has many parameters, most of them
 are held fixed when we fit the data for calibration purposes.  The
 continuum components are fixed and the interstellar absorption components
are held fixed.  All the line energies and widths are also held fixed.
Typically, we freeze all line normalizations except for the four
normalizations of the brightest lines/line complexes.  We allow the 
normalizations for the
\ion{O}{vii}~He$\alpha$~{\em r} line, the \ion{O}{viii}~Ly$\alpha$ line, 
the \ion{Ne}{ix}~He$\alpha$~{\em r} line, and \ion{Ne}{X}~Ly$\alpha$
line to vary in the fit.  For the \ion{O}{vii}~He$\alpha$ and 
\ion{Ne}{ix}~He$\alpha$ triplets, we link the normalizations of the
{\em f}, {\em i}, and {\em r} lines to each other and only allow one of them
to vary during the fitting process.  In this way, the triplet can
increase or decrease its normalization as a group but the
normalizations of the individual lines in the triplet can not vary
independently of each other.  There is also a global constant that
multiplies the entire spectrum that is allowed to vary.  In this
manner, we allow only 5 of the 208 parameters in the IACHEC model to
vary when we fit. We are essentially allowing the normalizations of the
four brightest line/line complexes to vary while freezing the weaker
lines and the continuum.  We assume that E0102 is
not changing significantly over the 19 year lifetime of the CXO
mission such that the flux from the source in 1999 is not
significantly different from the flux in 2018.  And therefore we
assume the total flux in a given line is not changing or changing very
little over the 19 year mission.
 
   \begin{figure}
   \begin{center}
   \begin{tabular}{c}
   \includegraphics[width=4.0in,angle=270]{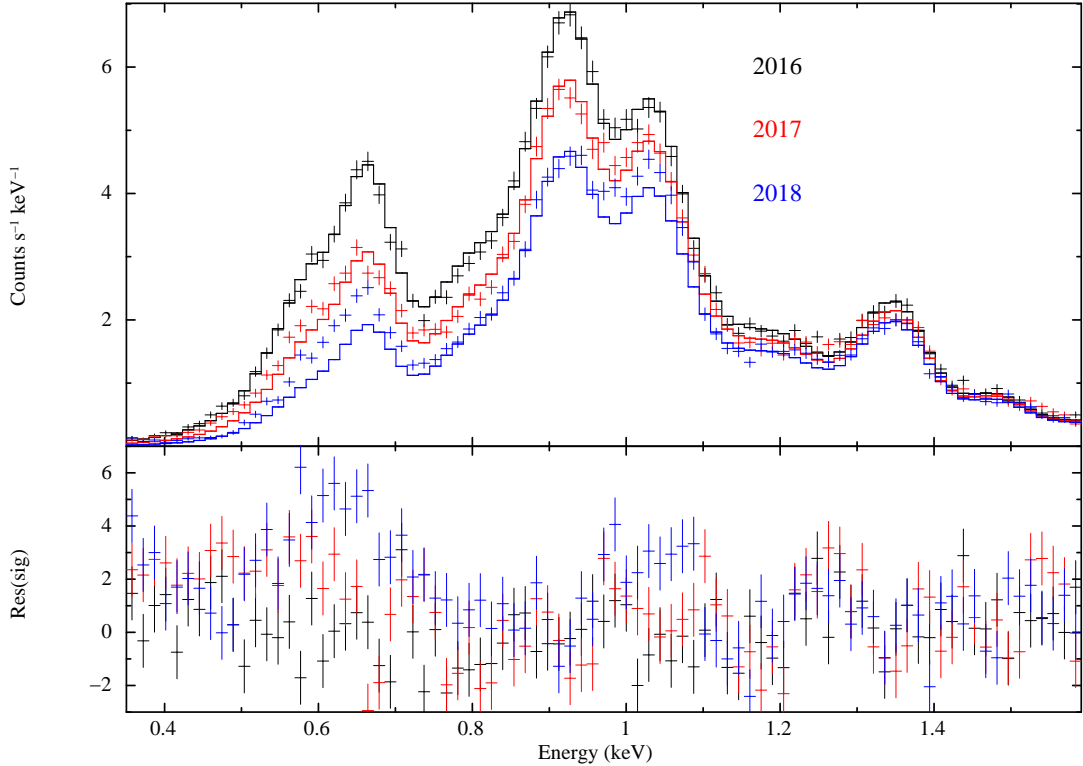}
   \end{tabular}
   \end{center}
   \caption[]
   { \label{fig:s3_e0102_sp} ACIS-S3 spectra of E0102 from OBSIDs
     18418(2016), 19850(2017)  and 20639(2018). The 2016 data are fit with the
     standard IACHEC model and that model is overplotted on the 2017
     and 2018 data.  

}
   \end{figure} 

\subsection{ACIS-S Results}
\label{sec:acis_s}

E0102 has been observed many times with ACIS-S since the beginning of
the mission. The mirrors on the CXO mission produce such sharp X-ray
images that observations of a point source can be affected by
``pileup''. ``Pileup'' is defined as two photons interacting with the CCD
within one detection cell (typically a $3\times3$ pixel region) within
one readout frame of the CCD.  Even though E0102 is an extended
source, some of the bright filaments in E0102
are bright enough to have significant pileup.  Most of the
observations of E0102 early in the mission were executed in
``full-frame'' mode with an integration time of 3.2~s.  We have
excluded the ``full-frame'' observations from our analysis and selected
only the ``subarray'' observations with shorter frametimes of 1.1~s,
0.8~s, and 0.4~s in order to minimize the effects of pileup on our
data. There are 32 subarray observations of E0102 on S3 included in
our analysis listed in Table~\ref{tab:s3obs}.  Most of these
observations are near the center of the CCD with {\tt chipy} values
around 512, but 13 of the observations are at different {\tt chipy}
positions.

   \begin{figure}
   \begin{center}
   \begin{tabular}{c}
   \includegraphics[width=5.5in]{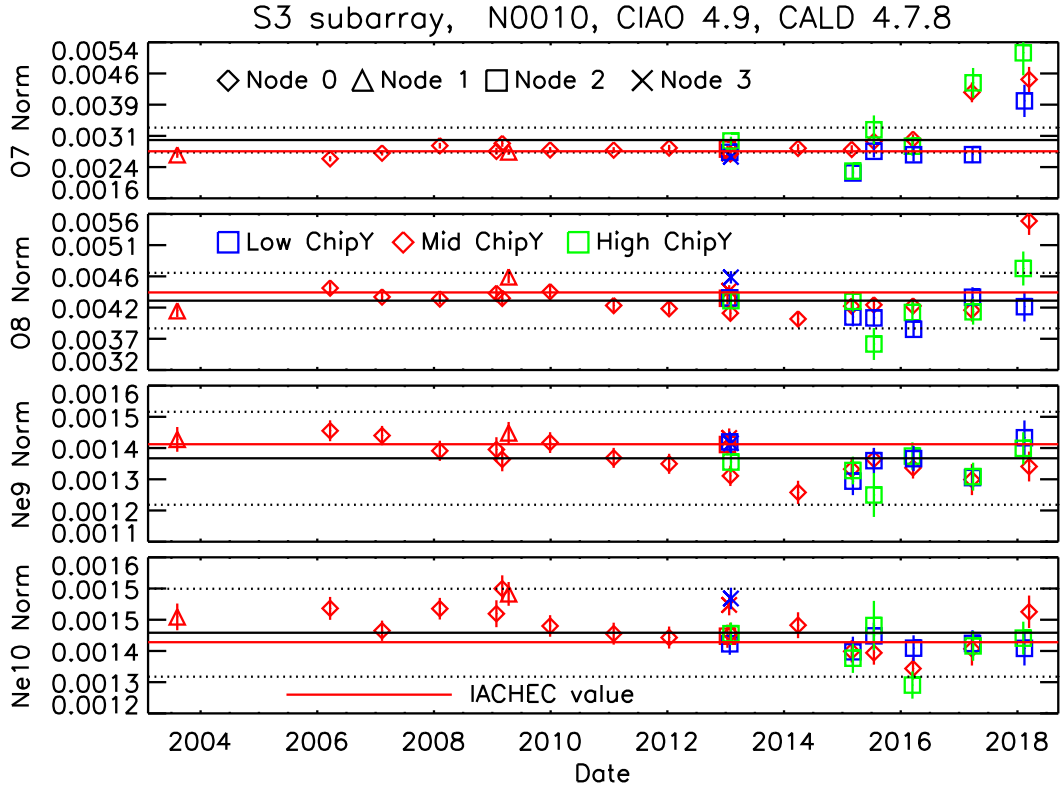}
   \end{tabular}
   \end{center}
   \caption[]
   { \label{fig:e0102S3} Line normalizations from E0102 on S3 as a
     function of time.  The solid black line is the average of the
     data points near the on-axis point aimpoint.  The red dashed
     lines are $+/-10\%$ above and below the average.  The points away
     from the nominal aimpoint are indicated in green and blue. 
}
   \end{figure} 

We have fit all of the S3 observations with the standard IACHEC model
allowing only the global normalization and the normalizations for the
\ion{O}{vii}~He$\alpha$~{\em r} line, the \ion{O}{viii}~Ly$\alpha$ line, 
the \ion{Ne}{ix}~He$\alpha$~{\em r} line, and \ion{Ne}{X}~Ly$\alpha$
line to vary. Figure~\ref{fig:s3_e0102_sp} shows an example of these fits
for the three most recent observations near the S3 aimpoint from 2016,
2017 and 2018.  The model was fit to the 2016  data and then frozen
for the 2017 and 2018 observations to demonstrate deficiencies in the 
time-dependent contamination model. The large residuals in the 2018
spectrum at the  \ion{O}{viii}~Ly$\alpha$ and  \ion{Ne}{X}~Ly$\alpha$
lines indicate that the contaminant is over-estimated.
Note that the difference between the 2017 and 2018 observations is not
as large as the model predicts.  The residuals
indicate that the \ion{O}{viii}~Ly$\alpha$ line and the 
\ion{Ne}{X}~Ly$\alpha$ are not well fitted in 2018 but the 
\ion{Ne}{ix}~He$\alpha$~{\em r} line is well fitted.  This will be
challenging to correct with a revised contamination model.

 We compared the fitted line normalizations for the
\ion{O}{vii}~He$\alpha$~{\em r} line, the \ion{O}{viii}~Ly$\alpha$
line, the \ion{Ne}{ix}~He$\alpha$~{\em r} line, and the \ion{Ne}{X}~Ly$\alpha$
line as a function of time.  The results are plotted in
Figure~\ref{fig:e0102S3}.  The solid black line is the average of the
on-axis data and the black dashed lines are +/-10\% from the
average. 
Figure~\ref{fig:e0102S3} shows that the line
normalizations are mostly consistent to within $\pm10\%$ from 2003
through 2016 for the on-axis data points with the exception of the
\ion{O}{vii}~He$\alpha$~{\em r} line.  After 2016, the 
\ion{O}{vii}~He$\alpha$~{\em r} line and \ion{O}{viii}~Ly$\alpha$
deviate dramatically from the previous values.  The 2018
normalizations on-axis for the \ion{O}{viii}~Ly$\alpha$ line
and \ion{O}{vii}~He$\alpha$~{\em r} line are
$\sim28\%$ and $\sim49\%$ higher than the average value.  
The \ion{Ne}{ix}~He$\alpha$~{\em r} line, and 
\ion{Ne}{X}~Ly$\alpha$ line normalizations are consistent with the
average within 10\%
 so the problem on S3 appears to effect energies below 0.9~keV.

\subsection{ACIS-I Results}   
\label{sec:acis_i}

There are 16 subarray observations of E0102 on the I3 CCD in the
ACIS-I array.  Table~\ref{tab:i3obs} lists the observations with their
locations on the CCD and exposure times and count rates.  Unlike the
S3 CCD where the aimpoint is near the middle of the CCD, the aimpoint
on the I3 CCD is near the top, right  corner (high {\tt chipx} and
{\tt chipy}).  Hence most of these observations have {\tt chipx} of
$\sim875$ and {\tt chipy} values of $\sim930$.  This position is close
to the center of the OBF-I filter, so the contamination layer is
thinner at this position than near the edges.  There are only 3 of the
14 observations that are at positions other than the nominal aimpoint.

\input{i3tab.tex}

\begin{figure}
   \begin{center}
   \begin{tabular}{c}
   \includegraphics[width=4.0in,angle=270]{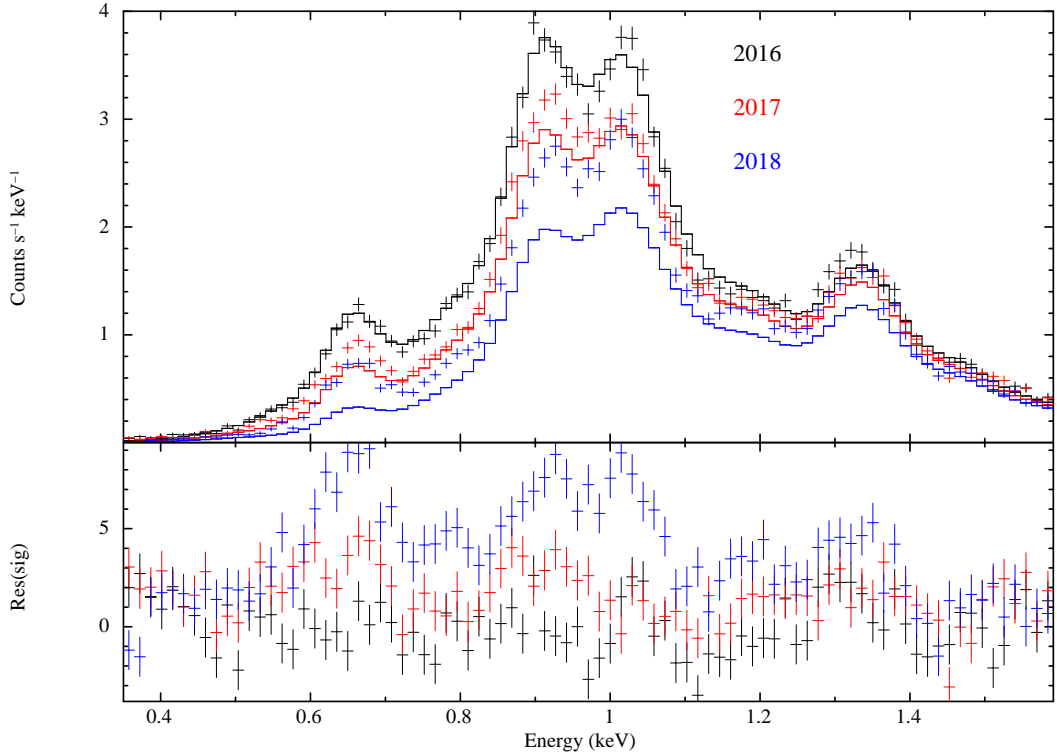}
   \end{tabular}
   \end{center}
   \caption[]
   { \label{fig:i3_e0102_sp} ACIS-I3 spectra of E0102 from OBSIDs
     18417(2016), 19849
     (2017)  and 20638(2018). The 2016 data are fit with the
     standard IACHEC model and that model if overplotted on the 2017
and 2018 data.  

}
   \end{figure}

Figure~\ref{fig:i3_e0102_sp} shows an example of these fits
for the three most recent observations near the I3 aimpoint from 2016,
2017 and 2018.  Again, the model was fit to the 2016  data and then frozen
for the 2017 and 2018 data to demonstrate deficiencies in the time-dependent
contamination model.  Note the dramatic
difference in the expected model spectrum for the 2018 data.
The N0010 contamination model is over-estimating the contamination by a
large amount at the aimpoint on I3.  This is partly due to the fact
that the accumulation rate has decreased but it is also due to
the fact that the N0010 contamination model predicted significantly
more contamination at the aimpoint on I3 than S3 (see
Figures~\ref{fig:akos_aciss} and ~\ref{fig:akos_acisi}).

\begin{figure}
   \begin{center}
   \begin{tabular}{c}
   \includegraphics[width=5.5in]{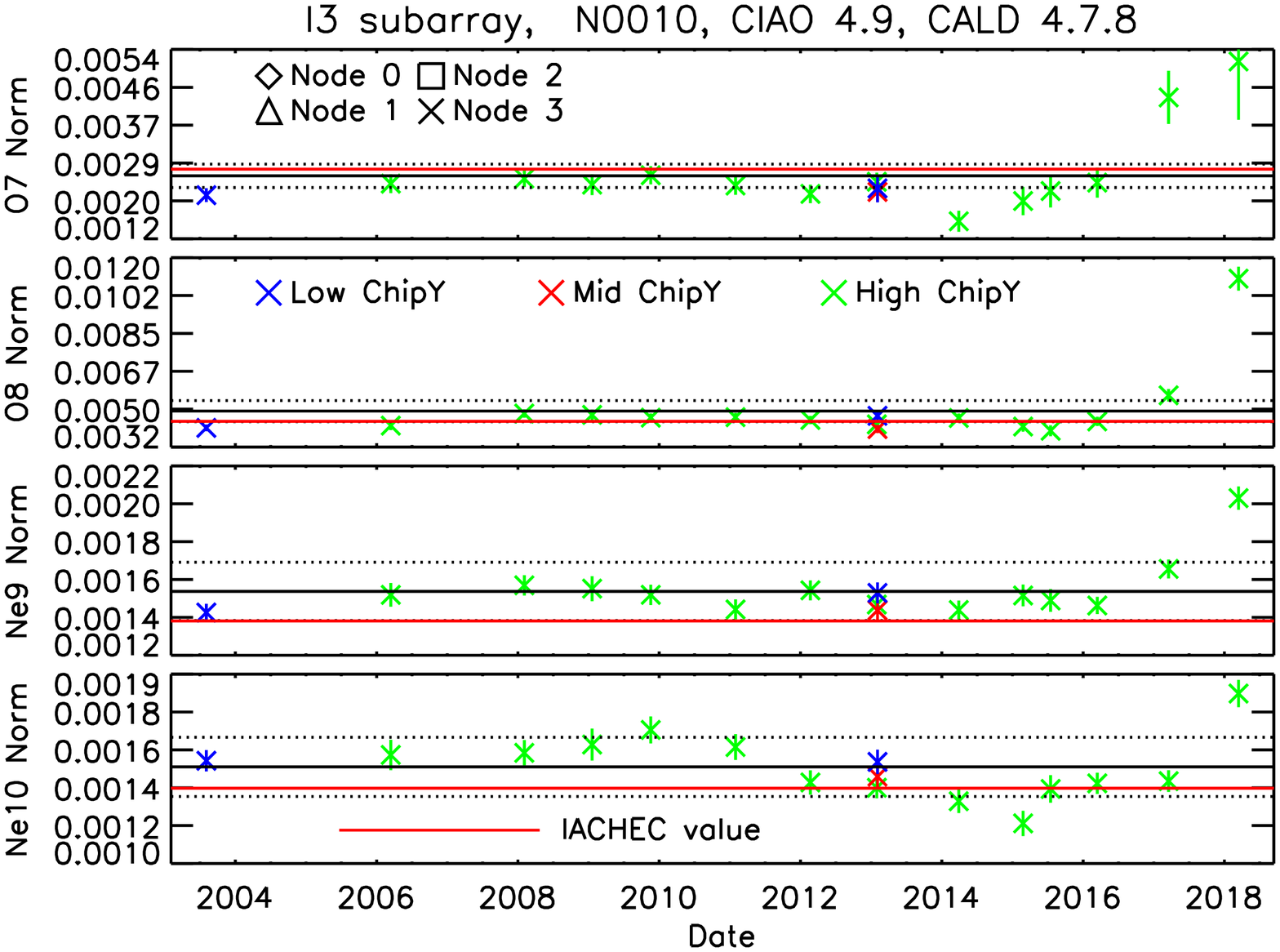}
   \end{tabular}
   \end{center}
   \caption[]
   { \label{fig:e0102I3} Line normalizations from E0102 on I3 as a
     function of time.  The solid black line is the average of the
     data points near the on-axis point aimpoint.  The red dashed
     lines are $+/-10\%$ above and below the average.  The points away
     from the nominal aimpoint are indicated in red and blue. 
}
   \end{figure}

 We compared the fitted line normalizations for the
\ion{O}{vii}~He$\alpha$~{\em r} line, the \ion{O}{viii}~Ly$\alpha$
line, the \ion{Ne}{ix}~He$\alpha$~{\em r} line, and the \ion{Ne}{X}~Ly$\alpha$
line as a function of time.  The results are plotted in
Figure~\ref{fig:e0102I3}.  The over-correction for the contamination
layer in 2018 is large.  The normalizations for the
\ion{O}{vii}~He$\alpha$~{\em r} line, the \ion{O}{viii}~Ly$\alpha$ line, the
\ion{Ne}{ix}~He$\alpha$~{\em r} line, and the
\ion{Ne}{X}~Ly$\alpha$ line
are over-estimated by  $\sim98\%$, $\sim125\%$, $\sim32\%$, and
$\sim25\%$ respectively.
The data from 2016 and earlier are mostly consistent with
each other to within 10\%.  The discrepancy begins in 2017 and dramatically worsens
in 2018.  The revised contamination file soon to be released by the
CXC should address most of this dicrepancy.

\section{Possible Explanations For the Reduction in the Accumulation Rate} 

\begin{figure}
   \begin{center}
   \begin{tabular}{c}
   \includegraphics[trim={0.7cm 0.5cm 1.3cm 0.5cm},clip,width=6.5in]{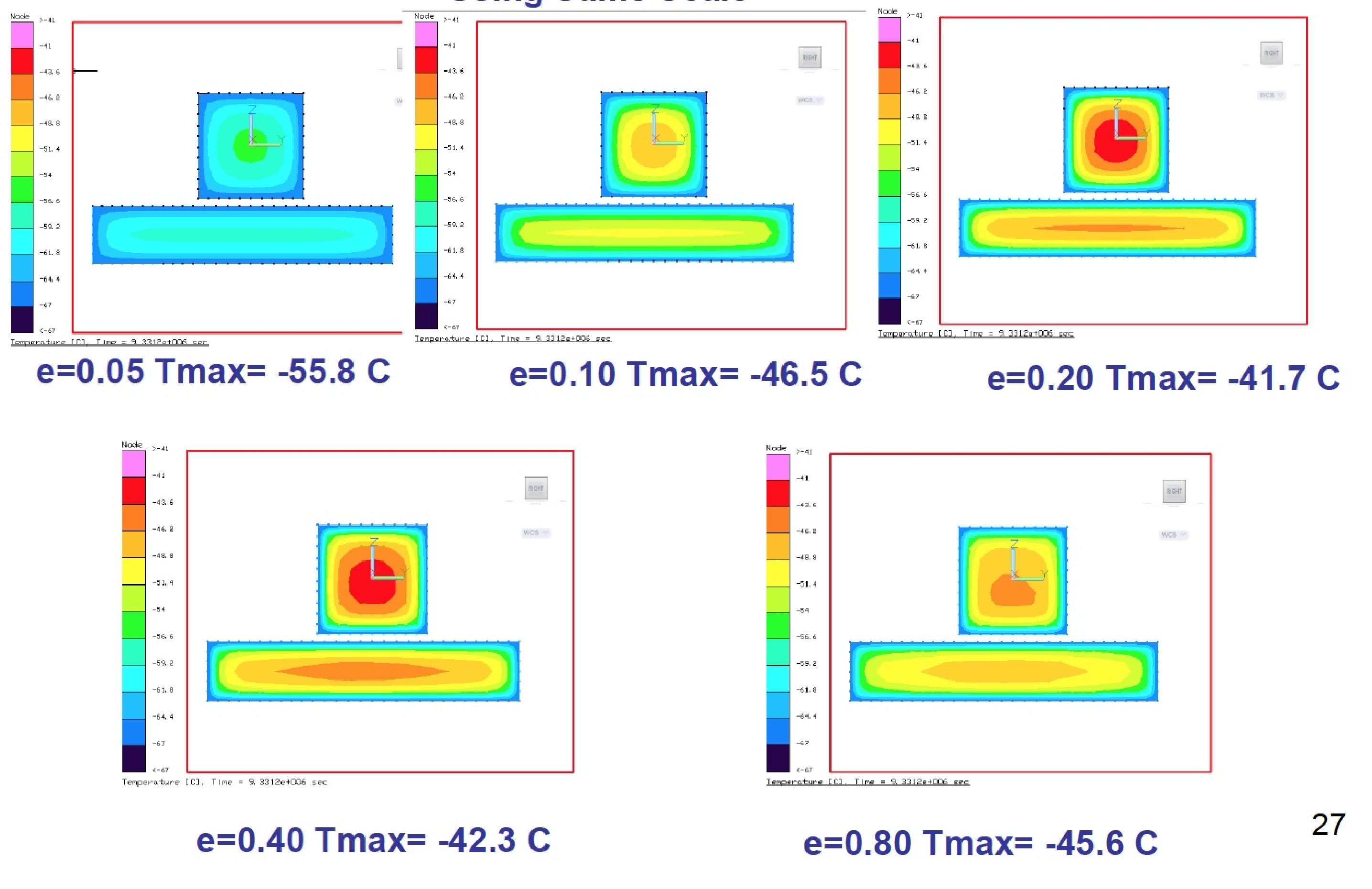}
   \end{tabular}
   \end{center}
   \caption[]
   { \label{fig:tice_emittance} The expected temperature distributions
     on the OBF-S and OBF-I as a function of emittance.

}
  \end{figure}

The analyses presented up to this point measure the accumulation rate
of the contaminant which is the difference between the deposition rate
and the vaporization rate.  If the accumulation changes, we do not
know if the deposition rate changed or the vaporization rate changed
or both.  Over the course of the mission, many components on the CXO
spacecraft have increased in temperature, reaching mission high
values within the last few years.  It is conceiveable that a component
on the spacecraft was not out-gassing significantly early in the
mission, but as its temperature increased it began to out-gas at a
higher rate.  Perhaps
the out-gassing from this component has now started to decrease, as the
source of the contaminant has diminished.
Another possibility is that the temperature distributions on the
filters have changed with time.  Figure~\ref{fig:tice_emittance} shows
the expected temperature distributions on the filters in the presence of
no contamination when the emittance is expected to be 0.05.  In this
case, the center of the OBF-I is at -55.8~C and the center of the
OBF-S is at $\sim -58.0$~C.  As contaminant accumulates on the filters
and the surrounding surfaces the temperature distribution will change,
with the centers of the filters becoming warmer.  For an emittance of
0.20, the center of the OBF-I increases to -41.7~C and the center of the
OBF-S increases to $\sim -46.0$~C.  The temperatures of the OBFs increase as
the emittance increases because the OBFs are more coupled to the
temperature of the warm optical bench assembly (+20~C). But as the
emittance continues increasing the OBF temperatures start to decrease
again because in this model, the surfaces around the OBF are also
accumulating a contamination layer and those surfaces have a higher
emittance which results in better coupling between those relatively
cold surfaces and the OBFs.

As shown previously\cite{odell2013}, the
vaporization rate of materials is a steep function of temperature with
the vaporization rate increasing by roughly one order of magnitude for
every 5~C increase in temperature. Therefore, it is possible that the
vaporization rate  has increased by about two to three orders of magnitude
in the centers of the filters as the temperatures have increased from
$\sim -56$~C to $\sim-42$~C. .  This could be part of the explanation
for the reduction in the accumulation rate that has been observed. This
would be consistent with the center of the OBF-I showing a larger
reduction in the accumulation rate than the center of the OBF-S since
the center of the OBF-I is warmer than the center of the OBF-S.

\section{Future Work}   
\label{sec:future}

  The temporal model of the contamination correction in the {\tt N0010} file
{\tt acisD1999-08-13contamN0010.fits} contained in CALDB~4.7.8 needs
modification to predict less absorption near the center and edges of the OBFs.
This is clear from the E0102 line normalizations presented in 
Figures~\ref{fig:e0102S3} and~\ref{fig:e0102I3}.  The CXC calibration
team is working on a
revision to the {\tt N0010} model that will change the time dependence
of the spatial distribution and will update the chemical composition
as a function of time. We expect this revised contamination model to
be released in two stages (both in 2018), one release for the OBF-I
and one for OBF-S.

The accumulation rate of the contaminant will need to be monitored
more frequently in the coming years. The accumulation rate at the
centers and edges of the OBFs for both ACIS-S and ACIS-I have all
changed in unexpected ways over the past two years.  The continued
characterization of these accumulation rates with time may provide 
constraints on the deposition and vaporization rates.

The CXO project considered a ``Bakeout'' of the ACIS
instrument\cite{plucinsky2004} soon after the contamination layer was
discovered in 2004.  The project decided at that time that a Bakeout
was not worth the risk.  There have been several papers written
describing models of an ACIS Bakeout, see
O'Dell~\etal~(2005)\cite{odell2005}, O'Dell~\etal~(2013)\cite{odell2013}, and 
O'Dell~\etal~(2015)\cite{odell2015}.  These papers predict a range of
outcomes from successful to unsuccessful depending on the assumed
volatilities for the contaminants.  The recently discovered reduction
in the accumulation rate makes it less likely the project will
consider a Bakeout worth the risk. Nevertheless, we will continue to
monitor the accumulation rate and spatial distribution of the contaminant
to constrain the volatilities of the possible contaminants to hopefully to
constrain the range of possible outcomes for a Bakeout. If the
contaminant were observed to decrease in the center of the OBF-I and
OBF-S, we would know that the vaporization rate is larger than the
deposition rate at the current temperatures. Such a result would
indicate that a Bakeout is likely to be successful, even at
temperatures not much higher than the current range of -70~C to -42~C.

\section{Conclusions} 
\label{sec:conclusions}

We have presented the accumulation rate of the ACIS contamination
layer as a function of time.  The accumulation rate decreased from launch
until 2005, was fairly linear from 2005 to 2010,  increased
after 2010 but has sharply decreased since 2016.  The chemical
composition of the contamination has changed
with time, possibly indicating that multiple sources are responsible
for the contamination.  The C, O, and F all exhibit different time
dependencies again indicating that multiple materials have accumulated
at different rates over the course of the mission. Nevertheless, all
three have shown a dramatic decrease over the past year.  The
explanation for this sudden decrease is not clear.  It could be that
the deposition rate has decreased or the vaporization rate has
increased, or both.  The CXC will need to monitor the contamination
layer frequently with dedicated calibration observations over the
coming years to accurately model the contamination layer

We tested the current contamination model {\tt N0010} with the SNR
E0102. We find that the fitted values for the normalizations of the 
\ion{O}{vii}~He$\alpha$~{\em r} line, the \ion{O}{viii}~Ly$\alpha$ line, 
the \ion{Ne}{ix}~He$\alpha$~{\em r} line, and \ion{Ne}{X}~Ly$\alpha$
line are mostly consistent to within $\pm10\%$ for both ACIS-S and
ACIS-I near the
aimpoint from 2003 through 2016.  After 2016, the line normalizations
begin to deviate from the average value, with deviations as large as
49\% at the aimpoint on ACIS-S and 125\% at the aimpoint on ACIS-I
for the \ion{O}{viii}~Ly$\alpha$ line. The CXC is preparing a revised 
contamination file
that will significantly improve the agreement from 2016 onwards for
release this year.

\acknowledgments 
We acknowledge support under NASA contract NAS8-03060. 
The {\em Chandra X-ray Observatory} is operated by the
Smithsonian Astrophysical Observatory under contract to the NASA
Marshall Space Flight Center (MSFC). The Advanced CCD Imaging
Spectrometer (ACIS) was developed by the Massachusetts Institute of
Technology and the Pennsylvania State University.\\
We sincerely thank all of our colleagues in the IACHEC that
contributed to the development of the highly successful spectral model
for E0102.

\bibliography{spie2018} 
\bibliographystyle{spiebib} 

\end{document}

%% file: s3tab.tex
\begin{table}[ht]
\caption{ACIS S3 Observations of E0102} 
\label{tab:s3obs}
\begin{center}       
\begin{tabular}{rcrrcrrcrc} 
\hline
ObsID	& Date	   & ChipX  & ChipY	& Node	& Exposure	& Counts	& Frame	& 1stRow & Nrows \\
	&	   & 	    &           &	& (s)		& (0.3-2keV)    & (s)   &         &  \\
\hline									
 3545	& 2003.60  & 357.6  & 496.8	& 1	& 7863.8	&   57133	&1.1	&385	&256	\\
 6765	& 2006.22  & 106.8  & 498.8	& 0	& 7636.7	&   51768	&0.8	&384	&256	\\
 8365	& 2007.11  & 100.2  & 494.3	& 0	&20985.3	&  138698	&0.8	&384	&256	\\
 9694	& 2008.10  & 100.1  & 492.8	& 0	&19196.5	&  124804	&0.8	&384	&256	\\
10654	& 2009.17  &  98.9  & 495.2	& 0	& 7307.1	&   45532	&0.8	&335	&256	\\
10655	& 2009.17  &  98.2  & 493.0	& 0	& 6810.7	&   43234	&0.4	&433	&128	\\
10656	& 2009.18  & 342.0  & 494.3	& 1	& 7763.9	&   48603	&0.8	&335	&256	\\
11957	& 2009.99  & 103.2  & 491.1	& 0	&18447.8	&  112437	&0.8	&335	&256	\\
13093	& 2011.08  &  89.8  & 487.9	& 0	&19049.4	&  108275	&0.8	&335	&256	\\
14258	& 2012.03  &  86.1  & 493.7	& 0	&19049.3	&  102049	&0.8	&360	&256	\\
15555	& 2013.03  & 676.5  & 489.5	& 2	&23837.3	&  114679	&0.8	&360	&256	\\
15558	& 2013.06  & 854.4  & 479.7	& 3	&23051.7	&  107679	&0.8	&360	&256	\\
15556	& 2013.07  & 666.1  & 164.2	& 2	&23841.3	&   94733	&0.8	& 42	&256	\\
15467	& 2013.08  &  91.7  & 488.9	& 0	&19082.2	&   92610	&0.8	&360	&256	\\
15557	& 2013.09  & 657.9  & 917.2	& 2	&24191.7	&   78972	&0.8	&769	&256	\\
15559	& 2013.09  & 854.8  & 162.4	& 3	&23842.1	&   92216	&0.8	& 42	&256	\\
16589	& 2014.24  &  78.1  & 485.5	& 0	& 9569.5	&   40194	&0.8	&360	&256	\\
17380	& 2015.16  & 119.2  & 489.0	& 0	&17655.1	&   65808	&0.8	&360	&256	\\
17381	& 2015.18  & 659.7  & 165.6	& 2	& 9573.6	&   27095	&0.8	& 42	&256	\\
17382	& 2015.18  & 671.5  & 926.7	& 2	& 9572.7	&   21161	&0.8	&769	&256	\\
17688	& 2015.54  & 110.7  & 481.6	& 0	& 9569.6	&   33973	&0.8	&360	&256	\\
17689	& 2015.54  & 661.3  & 158.0	& 2	& 9573.5	&   25482	&0.8	& 42	&256	\\
17690	& 2015.54  & 660.8  & 914.2	& 2	& 9572.7	&   20772	&0.8	&769	&256	\\
18418	& 2016.21  &  99.0  & 480.2	& 0	&14326.2	&   45687	&0.8	&360	&256	\\
18420	& 2016.20  & 651.9  & 919.2	& 2	&19085.4	&   36556	&0.8	&769	&256	\\
18419   & 2016.22  & 642.3  & 157.9     & 2     &19084.7        &  45491 	&0.8    & 42    &256     \\
19850   & 2017.22  &  98.6  & 484.9     & 0     &14326.21       &  38782        &0.8    &360    &256     \\
19851   & 2017.23  & 647.6  & 166.6     & 2     &19085.35       &  38603        &0.8    & 42    &256     \\
19852   & 2017.24  & 652.3  & 922.4     & 2     &19085.38       &  30461        &0.8    &769    &256     \\
20639   & 2018.20  & 103.8  & 489.4     & 0     &14326.21       &  32681        &0.8    &360    &256     \\
20640   & 2018.12  & 645.2  & 165.3     & 2     &19013.47       &  32687        &0.8    & 42    &256     \\
20641   & 2018.10  & 658.2  & 923.1     & 2     &19084.61       &  26226        &0.8    &769    &256     \\
\hline 
\end{tabular}
\end{center}
\end{table}

%% file: i3tab.tex
\begin{table}[ht]
\caption{ACIS I3 Observations of E0102} 
\label{tab:i3obs}
\begin{center}       
\begin{tabular}{rcrrcrrcrc} 
\hline
ObsID	& Date	   & ChipX  & ChipY	& Node	& Exposure	& Counts	& Frame	& 1stRow & Nrows \\
	&	   & 	    &           &	& (s)		& (0.3-2keV)    & (s)   &         &  \\
\hline									
 3526	& 2003.59  & 884.6  & 120.8	& 3	 & 7859.9	 & 21368	& 1.0	& 1	& 256  \\	
 6756	& 2006.20  & 888.7  & 922.5	& 3	 & 7159.9	 & 23197	& 0.8	& 768	& 256	\\
 9690	& 2008.09  & 884.4  & 925.6	& 3	 & 19192.5	 & 60496	& 0.8	& 768	& 256	\\
10649	& 2009.05  & 880.3  & 926.4	& 3	 & 7637.5	 & 23510	& 0.8	& 768	& 256	\\
11956	& 2009.98  & 882.6  & 922.3	& 3	 & 19189.4	 & 57787	& 0.8	& 768	& 256	\\
13092	& 2011.08  & 875.1  & 940.0	& 3	 & 19050.1	 & 53984	& 0.8	& 768	& 256	\\
14257	& 2012.04  & 881.9  & 928.7	& 3	 & 19051.0	 & 51539	& 0.8	& 768	& 256	\\
15466	& 2013.08  & 875.5  & 937.8	& 3	 & 19082.9	 & 47984	& 0.8	& 768	& 256	\\
15471	& 2013.09  & 877.4  & 505.8	& 3	 & 9574.3	 & 22301	& 0.8	& 341	& 256	\\
15472	& 2013.09  & 873.9  & 144.6	& 3	 & 9574.3	 & 16405	& 0.8	&  1	& 256	\\
16588	& 2014.24  & 875.4  & 953.9	& 3	 & 9571.9	 & 21634	& 0.8	& 768	& 256	\\
17379	& 2015.15  & 872.9  & 929.4	& 3	 & 17616.6	 & 35717	& 0.8	& 768	& 256	\\
17687	& 2015.54  & 866.5  & 934.0	& 3	 & 13285.4	 & 25598	& 0.8	& 768	& 256	\\
18417	& 2016.20  & 870.5  & 938.8	& 3	 & 22886.1	 & 40565	& 0.8	& 768	& 256	\\
19849   & 2017.21  & 877.0  & 938.6     & 3      & 23837.4       & 36761        & 0.8   & 768   & 256   \\    
20638   & 2018.20  & 877.7  & 933.7     & 3      & 23838.1       & 32164        & 0.8   & 768   & 256   \\
\hline 
\end{tabular}
\end{center}
\end{table}